\documentclass[final,3p,times,twocolumn]{elsarticle}
\usepackage{algorithm}
\usepackage{algpseudocode}
\usepackage{lineno}
\modulolinenumbers[5]

\journal{Structural and Multidisciplinary Optimization}
\usepackage[table,xcdraw]{xcolor}
\usepackage{graphicx}
\usepackage{dcolumn}
\usepackage{bm}
\usepackage{multirow} 
\usepackage{color}
\usepackage{graphicx} 
\usepackage{mathtools}
\usepackage{multirow}
\usepackage{caption}
\usepackage{amsmath} 
\usepackage{chemformula} 
\usepackage{csquotes} 
\biboptions{numbers,sort&compress}
\usepackage{subfig}
\usepackage{placeins}
\usepackage{cancel}
\usepackage[normalem]{ulem}
\usepackage{siunitx}
\usepackage{bm}
\usepackage[capitalise]{cleveref}
\usepackage{xr}
\usepackage{comment}
\usepackage{csvsimple}
\usepackage{breqn}
\usepackage{amsmath}
\usepackage{array} 
\newcolumntype{P}[1]{>{\centering\arraybackslash}p{#1}}

\makeatletter
\newcommand*{\addFileDependency}[1]{
  \typeout{(#1)}
  \@addtofilelist{#1}
  \IfFileExists{#1}{}{\typeout{No file #1.}}
}
\makeatother

\begin{document}

\begin{frontmatter}
\title{Transformer-based Topology Optimization}

\author[A]{Aaron Lutheran}
\author[B]{Srijan Das}
\author[A,C]{Alireza Tabarraei\corref{mycorrespondingauthor}}
\cortext[mycorrespondingauthor]{Corresponding author}
\ead{atabarra@charlotte.edu}

\address[A]{Department of Mechanical Engineering and Engineering Science, The University of North Carolina at Charlotte, Charlotte, NC 28223, USA}
\address[B]{Department of Computer Science, The University of North Carolina at Charlotte, Charlotte, NC 28223, USA}
\address[C]{School of Data Science, The University of North Carolina at Charlotte, Charlotte, NC 28223, USA}

\begin{abstract}
Topology optimization enables the design of highly efficient and complex structures, but conventional iterative methods, such as SIMP-based approaches, often suffer from high computational costs and sensitivity to initial conditions. Although machine learning methods have recently shown promise for accelerating topology generation, existing models either remain iterative or struggle to match ground-truth performance. In this work, we propose a transformer-based machine learning model for topology optimization that embeds critical boundary and loading conditions directly into the tokenized domain representation via a class token mechanism. We implement this model on static and dynamic datasets, using transfer learning and FFT encoding of dynamic loads to improve our performance on the dynamic dataset. Auxiliary loss functions are introduced to promote the realism and manufacturability of the generated designs. We conduct a comprehensive evaluation of the model’s performance, including compliance error, volume fraction error, floating material percentage, and load discrepancy error, and benchmark it against state-of-the-art non-iterative and iterative generative models. Our results demonstrate that the proposed model approaches the fidelity of diffusion-based models while remaining iteration-free, offering a significant step toward real-time, high-fidelity topology generation.
\end{abstract}

\end{frontmatter}

\section{Introduction}\label{introduction}

Topology optimization is a foundational component of modern design, allowing engineers to create efficient and complex structural designs that conventional design methodologies often fail to achieve \cite{bendsoe2003topology, sigmund_topology_2013, deaton2014survey}. By optimizing the distribution of material within a design space, topology optimization algorithms can maximize the performance of a design for a given engineering objective, such as minimizing compliance, maximizing stiffness-to-weight ratio, or improving functional efficiency. The development of efficient, robust, and scalable topology optimization tools is therefore critical for advancing structural design in aerospace, automotive, civil, and biomedical applications, as well as additive manufacturing, where geometrical freedom is essential for lightweight yet reliable structures \cite{Zhu2021Review,  Meng2019Roadmap}.

Conventional algorithms for topology optimization are predominantly based on iterative gradient-driven methods. Among these, the solid isotropic material with penalization (SIMP) method and its extensions remain the most widely adopted formulations \cite{bendsoe1989optimal, yang1996stress, sigmund_topology_2013}. These algorithms alternate between finite element analysis (FEA), sensitivity computation, and design updates using mathematical programming techniques such as the method of moving asymptotes (MMA) \cite{svanberg1987mma}. While such methods are mathematically rigorous and capable of achieving high-quality designs, they are computationally demanding. Each iteration requires solving large-scale equilibrium equations, and the number of iterations needed for convergence increases significantly with problem size, mesh resolution, or the presence of multiple loading conditions. Furthermore, the iterative nature of these schemes renders them sensitive to initial guesses and heuristic parameter choices (e.g., penalization factors, filtering strategies, and continuation schemes). As a result, achieving designs with reliable convergence often requires multiple restarts, and real-time design generation, critical for rapid prototyping and interactive CAD workflows, remains out of reach \cite{mukherjee_accelerating_2021, behzadi_real-time_2021}.

To address these limitations, several acceleration strategies have been proposed. Parallel and distributed computing approaches decompose large domains to exploit high-performance computing resources \cite{borrvall_large-scale_2001}. Reduced-order modeling has also been employed to approximate structural response with fewer degrees of freedom, thereby reducing computational cost while preserving accuracy \cite{guest_reducing_2010, filomeno_coelho_model_2008}. Other efforts include multiresolution topology optimization, level-set methods, and isogeometric analysis, which aim to balance accuracy and efficiency for large-scale or high-fidelity applications \cite{guo2014doing, allaire2004structural}. Despite these advancements, the reliance on iterative solvers continues to impose high computational overhead, motivating a shift toward data-driven and surrogate-assisted methods.

In recent years, machine learning (ML) has shown considerable promise for accelerating topology optimization.  Examples include topology optimization using neural networks  \cite{chandrasekhar_tounn_2021, shishir_multimaterials_2024, tabarraei2025graph}, physics-informed neural networks \cite{jeong_physics-informed_2023}, and convolutional encoder–decoder frameworks that exploit spatial locality \cite{oh_deep_2019, senhora2022machine}. Generative adversarial networks (GANs) have also been employed to produce high-resolution designs \cite{nie_topologygan_2021}, while large-scale multi-grid learning strategies have demonstrated the potential of training with tens of thousands of 3D samples to achieve visually convincing designs \cite{rade2023deep}. More recently, quantum computing has also been explored in this context, including quantum annealing approaches for structural topology optimization \cite{wang2024mapping, ye2023quantum} and hybrid quantum machine learning methods \cite{tabarraei2025variational} where variational quantum circuits generate latent encodings that are decoded into optimized material distributions.



Iterative ML-based generative models, particularly diffusion-based methods, have also been proposed to overcome the limitations of direct predictors. Diffusion models such as TopoDiff \cite{maze_diffusion_2022} and subsequent variants \cite{giannone_diffusing_2023, zhang_research_2025, lutheran2025latent} leverage denoising processes guided by learned compliance or feasibility predictors to generate designs with exceptional structural fidelity. These methods achieve state-of-the-art performance on compliance error and constraint adherence but require hundreds of sampling steps at inference, reintroducing the computational burdens that TO aims to overcome.

Transformers provide a promising alternative foundation for non-iterative topology optimization. Initially developed for natural language processing \cite{vaswani2023attentionneed}, transformers have achieved breakthroughs in computer vision through the Vision Transformer (ViT) \cite{dosovitskiy2020image, zhai2022scalingvisiontransformers}. By representing input data as sequences of tokens and leveraging multi-head self-attention, transformers capture both local and long-range spatial dependencies without hand-crafted convolutional kernels \cite{he2015deepresiduallearningimage}. This characteristic is particularly relevant for structural optimization problems, where nonlocal interactions between distant loads and supports govern the global response of the structure. Early studies have begun to explore transformers for scientific computing and PDE-based learning tasks \cite{cao2021choose}, yet their integration into topology optimization remains limited.

In this work, we propose a transformer-based, iteration-free generative model for topology optimization that addresses the limitations of classical iterative solvers and mitigates the inference cost of diffusion models. Our framework embeds boundary conditions, loading information, and volume fraction into the tokenized representation through a class token, while processing physics-informed surrogate fields of stress and strain energy density via a Vision Transformer. Auxiliary loss functions are introduced to enforce physical constraints and manufacturability. Furthermore, we extend the approach to dynamic loading conditions by incorporating frequency-domain load features through FFT-based encoding and leveraging transfer learning from large static datasets. Taken together, this study contributes to the growing landscape of AI- and quantum-inspired computational design methods, offering a scalable, generalizable, and non-iterative framework that advances the pursuit of real-time, high-fidelity topology optimization.

\section{Topology Optimization}\label{sec:02}
Topology optimization is the algorithmic design process that seeks to determine the most efficient material distribution in a given domain. This design is dependent on a set of prescribed loads and boundary conditions, as well as what performance objective is under scrutiny. Often, the primary goal is to minimize compliance (for structural optimization problems) for a specific volume fraction of material. Other objectives, such as maximizing thermal transfer or controlling surrounding fluid flow, are also common applications for topology optimization.

\subsection{Static Structural Optimization}\label{static_topopt}

One of the most widely used formulations of the topology optimization problem is the SIMP method, in which the domain is discretized into finite elements \cite{yang1996stress, sigmund_topology_2013}. In a static structural optimization problem, the static equilibrium equation must be satisfied. It enforces that internal elastic forces balance the applied external loads of the structure. This equilibrium can be expressed as

\begin{equation}
\mathbf{Ku} = \mathbf{f} \label{Static Eq}\\
\end{equation}
where $\mathbf{K}$ is the global stiffness matrix, $\mathbf{u}$ is the global displacements vector, and $\mathbf{f}$ is the global force vector. The stiffness matrix encodes both the material properties and connectivity of the elements. Any proposed design $\rho$ influences the values within the global stiffness matrix, which reflects how the design influences the structural response of the mesh. In a structural optimization problem, we consider the compliance energy $c$ to be our optimization objective. The optimization problem is then formulated as

\begin{eqnarray}
\min_{\rho} C(\rho) = \mathbf{u}^T \mathbf{K}\mathbf{u}\label{Compliance}\\
V(\rho)/V_0 \leq f \label{Vol Frac}\\
0 \leq \rho_e \leq 1 \label{Clamp}
\end{eqnarray}
where $f$ is the prescribed volume fraction, $V_0$ is the volume of the design space, $V(\rho)$ is the volume of the design $\rho$, and $\rho_e$ is the density of a specific element. This formulates the compliance $C(\rho)$ as a function of the design and as an objective to minimize by controlling the design vector $\rho$. In this formulation, the compliance is twice the total strain energy of the system. Minimizing the compliance corresponds to maximizing the overall stiffness, so a solution to this optimization problem will produce a structure with optimized stiffness.

The second line of the optimization problem arises from material usage. This constraint prevents the trivial case where all elements are fully solid, ensuring optimized designs have the desired volume fraction. In practical applications, this may represent a weight or material limit for a particular design problem. In addition to the global volume constraints, each design variable must be physically meaningful. The density is bounded between 0 and 1 to represent the minimum and maximum material amounts.

The stiffness $E_e$ of each element $e$ is used to construct the global stiffness matrix $K$ and is dependent on the element design variable $\rho_e$. For minimum finding algorithms, we need to keep continuous gradients and continuous design variables. This also means there needs to be a method for interpolating between the material properties of solid and void states. The effective Young's Modulus of each element in the SIMP method is given by

\begin{eqnarray}
    E_e(\rho_e) = E_{min} + \rho_e^p (E_0-E_{min})
\end{eqnarray}
$E_{min}$ is a minimum stiffness value to ensure the static equilibrium problem remains well defined for any valid structure. $p$ is the penalty factor, commonly set to 3, which serves to penalize intermediate densities between pure void $(\rho_e =0)$ and pure material $(\rho_e=1)$. 

The optimization process then updates the element densities based on the sensitivities of the compliance with respect to the design variables. These gradients become
\begin{eqnarray}
    \frac{\partial C}{\partial \rho_e} = -p\rho_e^{p-1}u^T_e\mathbf{K}_eu_e
\end{eqnarray}
where $u_e^T$ and $K_e$ are the element displacements and element stiffness matrices, respectively. These sensitivities quantify how local material adjustments affect global structural performance and form the foundation for gradient-based optimization updates. As such, it can be used in an optimization algorithm such as the Method of Moving Asymptotes (MMA) to update the design variables.

When the compliance has converged to a local minimum value, the optimization can be stopped. Stopping conditions are defined either by relative change in compliance or by iteration count limits. For the relative compliance stopping criteria, the condition is usually formulated as

\begin{equation}
        \frac{|C^{k+1}-C^k|}{C^k} < \epsilon
\end{equation}
where $\epsilon$ is a tolerance parameter that ensures subsequent iterations produce a small change in the compliance. The final state of the design vector $\rho$ is then considered the final optimized topology.

\begin{figure*}
    \centering
    \includegraphics[width=1\linewidth]{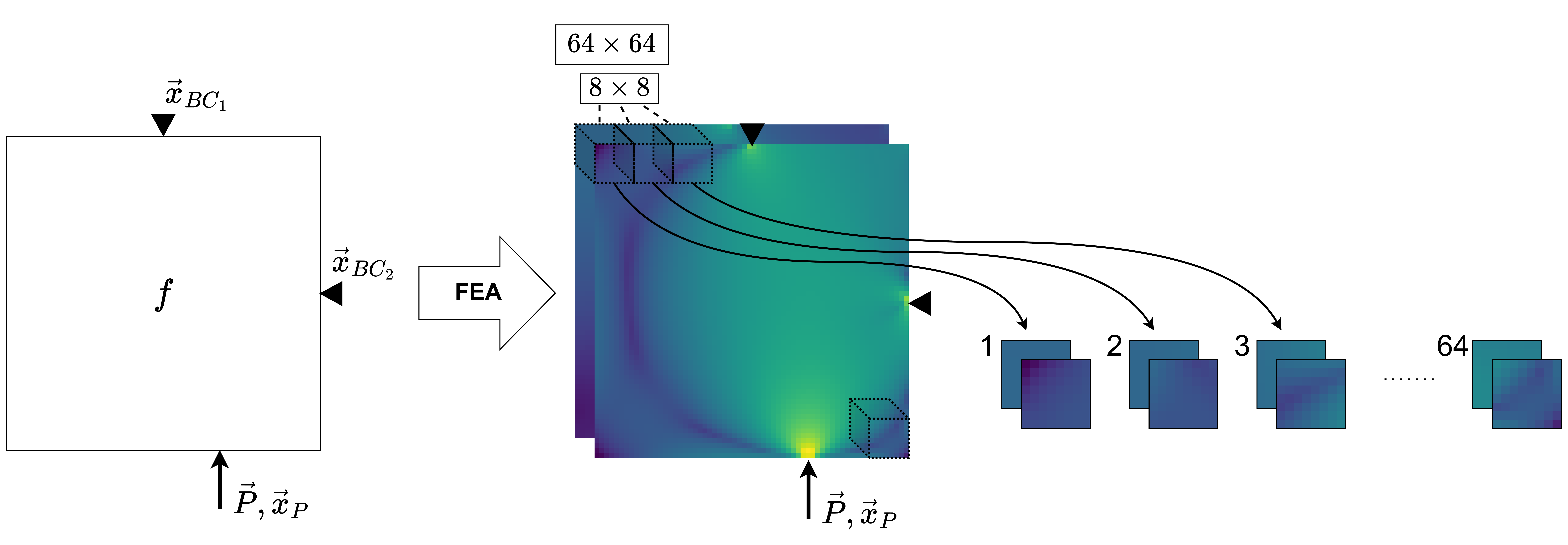}
    \caption{Processing for the inputs of the structural optimization problem for a given problem with boundary conditions $\vec{x}_{BC}$, load $\vec{P}$ at location $\vec{x}_P$, and volume fraction $f$. First, finite element analysis is used to convert the sparse inputs into dense images of the strain energy density and the von Mises stress. The fields of size $64 \times 64$ are then divided into a sequence of $64$ total $8 \times 8$ patches.}
    \label{fig:patches}
\end{figure*}

\subsection{Dynamic Structural Optimization}\label{dynamic_topopt}

The dynamic response of a structure under time-varying loads is governed by the finite element equation of motion, which incorporates inertia, damping, stiffness, and external forcing effects into the design's structural response \cite{lee2018efficient, rong2000topology, zhao2016dynamic, lee2015nonlinear}. For dynamic loading, the optimal topology is not necessarily captured by a static equivalent load, so a dynamic optimization method is necessary. This dynamic response can be expressed as

\begin{equation}
\mathbf{M} \ddot{\mathbf{u}}(t) + \mathbf{C} \dot{\mathbf{u}}(t) + \mathbf{K} \mathbf{u}(t) = \mathbf{f}(t)
\end{equation}
where the global mass matrix $\mathbf{M}(\rho)$, damping matrix $\mathbf{C}(\rho)$, and stiffness matrix $\mathbf{K}(\rho)$ are all functions of the design variable field $\rho$. The vector $\mathbf{f}(t)$ represents external time-dependent loading, while the displacement vector $\mathbf{u}(t)$ evolves according to the system dynamics. For each finite element $e$, the consistent mass formulation is obtained through integration of shape functions over the element domain 

\begin{equation}
\mathbf{M}_e = \int_{\Omega_e} \rho_e \, \mathbf{N}^T \mathbf{N} \, d\Omega
\end{equation}

This formulation ensures that inertia is distributed consistently with the displacement field approximation. In practice, however, lumped mass approximations are often employed for computational efficiency. In this case, the element mass matrix reduces to a diagonal form

\begin{equation}
\mathbf{M}_e^{\text{lumped}} = m_e \mathbf{I}
\end{equation}
with $m_e$ being the total element mass and $\mathbf{I}$ an identity mapping over nodal degrees of freedom. While less accurate in some contexts, lumped mass models significantly reduce computational cost in large-scale optimization problems. 

Damping effects are typically modeled using Rayleigh damping, which assumes that damping can be expressed as a linear combination of mass and stiffness contributions

\begin{equation}
\mathbf{C} = \alpha \mathbf{M} + \beta \mathbf{K}
\end{equation}
where coefficients $\alpha$ and $\beta$ are tuned to approximate desired modal damping ratios across relevant frequency ranges. This proportional damping model provides a balance between fidelity and computational tractability in optimization settings.  Mapping material properties to design variables requires interpolation schemes to ensure smooth transitions between solid and void states while avoiding numerical instabilities. Damping properties can be considered design-dependent, interpolated as

\begin{equation}
c(\rho_e) = c_{\min} + \rho_e^p (c_0 - c_{\min})
\end{equation}

The elastic modulus interpolation follows directly from the static compliance formulation and remains coupled to density penalization strategies such as SIMP (Solid Isotropic Material with Penalization).

The objective function in dynamic structural optimization often involves minimizing dynamic compliance under prescribed excitations. In the time domain, this objective is defined as

\begin{equation}
C_{\text{dyn}}(\rho) = \int_0^T \mathbf{f}(t)^T \mathbf{u}(t)\, dt \label{eqn:dyn_comp}
\end{equation}
which integrates work done by external forces over time. Sensitivity analysis provides gradients necessary for iterative optimization updates. In the time domain, derivatives of dynamic compliance with respect to design variables take the form

\begin{equation}
\frac{\partial C_{\text{dyn}}}{\partial \rho_e} = 
    \int_0^T 
    \mathbf{u}_e(t)^T 
    \frac{\partial \mathbf{K}_e}{\partial \rho_e}
    \,\mathbf{u}_e(t)\,
    dt \label{eqn:dyn_sens}
\end{equation}
which mirrors static sensitivity expressions but now integrates over transient responses. To solve these functions, we also need to discretize the problem again, in the time domain, rather than just the spatial discretization we performed with finite elements. Equations \cref{eqn:dyn_comp} and \cref{eqn:dyn_sens} become summations over $N$ discrete time-steps

\begin{align}
C_{\text{dyn}}(\rho) \approx \sum_{i=0}^N  \mathbf{f}(t_i)^T \mathbf{u}(t_i)\ \Delta t \label{eqn:dyn_comp_disc}\\
\frac{\partial C_{\text{dyn}}}{\partial \rho_e} \approx 
    \sum_{i=0}^N 
    \mathbf{u}_e(t_i)^T 
    \frac{\partial \mathbf{K}_e}{\partial \rho_e}
    \,\mathbf{u}_e(t_i)\,
    \Delta t \label{eqn:dyn_sens_disc}
\end{align}

Similar to the static optimization case, the MMA can still be used to update the design objectives due to the objective function remaining continuous. The same stopping conditions can also be implemented, operating on the dynamic compliance rather than the static compliance. This setup requires solving a system of linear equations, updating the positions and velocities, and calculating the element compliance sensitivities, all performed at every timestep. Once the dynamic response has been determined, the design can then be updated a single time, just to repeat the response and sensitivity calculations again. Increasing the mesh resolution or reducing the timestep $\Delta t$ has outsized effects on the overall runtime of the optimization scheme. Creating a system that could avoid this repeated discretization and produce an optimized topology without an iterative solver has massive potential to improve the dynamic topology optimization process.

\section{Transformers}\label{sec:3}
Transformer architectures have been used across many applications, such as sequence modeling for natural language processing and spatial modeling for image processing. Unlike convolutional neural networks, which rely on learning local relationships in image data \cite{he2015deepresiduallearningimage}, transformers use a self-attention mechanism to model the interactions between all input elements of a sample \cite{vaswani2023attentionneed}. This allows them to model non-local relationships, which greatly improves their performance on problems that cannot be assumed to have exclusively local behavior. Transformers are therefore suited for problems that relate distant parts of the input domain, such as the distant load and boundary conditions seen in topology optimization problems.

In order to leverage transformer models for structural optimization, we first need to represent the input conditions as a sequence. In image processing, a common technique for implementing transformers is to patchify the input image. The image is divided into non-overlapping image patches of size $P \times P$. This patching process converts the spatial domain into a collection of smaller, fixed-size regions that can be treated as an individual input unit, or \textit{token}, for use in a transformer model. An image of height $H$ and width $W$, the total number of patches $N$ is computed by $N = \frac{HW}{P^2}$. \cref{fig:patches} illustrates this process, demonstrating how the inputs for our structural optimization process are transformed into patches.

Each patch is then flattened into a vector and linearly projected into a token embedding space. Each flattened patch $x_i$ is multiplied by a learnable projection matrix $E \in \mathbb{R}^{D \times P^2}$ where $D$ is the embedded dimension. This results in an embedding vector $z_i^0$ for each patch with index $i \in \{1, 2, ... N\}$.  Since the same projection is applied to every patch, the resulting tokens lose information about their original locations in the input. Because spatial relationships are important for the final design, this positional information must be reintroduced into the tokens. A simple method is to prepend a position index to the new representation based on its position in the full image.

Transformers utilize a multi-head self-attention mechanism, which weighs the importance of each input token relative to all other input tokens. For a sequence of token embeddings $X \in \mathbb{R}^{N\times D}$, where $N$ is the number of tokens and $D$ is the length of the tokens, the self-attention mechanism performs three primary projections to the data

\begin{eqnarray}
    Q = XW^Q; \qquad K = XW^K; \qquad V = XW^V 
\end{eqnarray}

This projection maps the input token sequence $X$ to three mappings: the query $Q$, key $K$, and value $V$. This is a learnable projection via the weight matrices $W^Q$, $W^K$, and $W^V$, where each projection matrix lies in $\mathbb{R}^{D\times D}$. These projections are used for the computation of attention scores with scaled dot-product attention

\begin{equation}
    A_{ij} = \frac{\mathbf{q}_i^T \mathbf{k}_j}{\sqrt{d_k}}
\end{equation}
where $d_k$ is the dimensionality of the key vectors. The resulting attention values are normalized by a softmax function across all tokens

\begin{equation}
    A_{ij} =  \frac{\exp(A_{ij})}{\sum_{j=1}^N \exp(A_{ij})}
\end{equation}

These normalized weights quantify how strongly each token $i$ attends to token $j$. The updated representation for each token is then obtained as a weighted sum over the value vectors

\begin{equation}
    \mathbf{z}'_i = \sum_{j=1}^N A_{ij}\,\mathbf{v}_j
\end{equation}

This operation allows information from every token to influence representations in every other token, capturing long-range spatial dependencies across the image domain. Each token then passes through a feed-forward multilayer perceptron (MLP) block, where each layer applies a linear projection, a shift by a bias, and a nonlinear activation function. A residual connection is also used to stabilize training and preserve gradient flow.

\begin{figure}
    \centering
    \includegraphics[clip, width=1\linewidth]{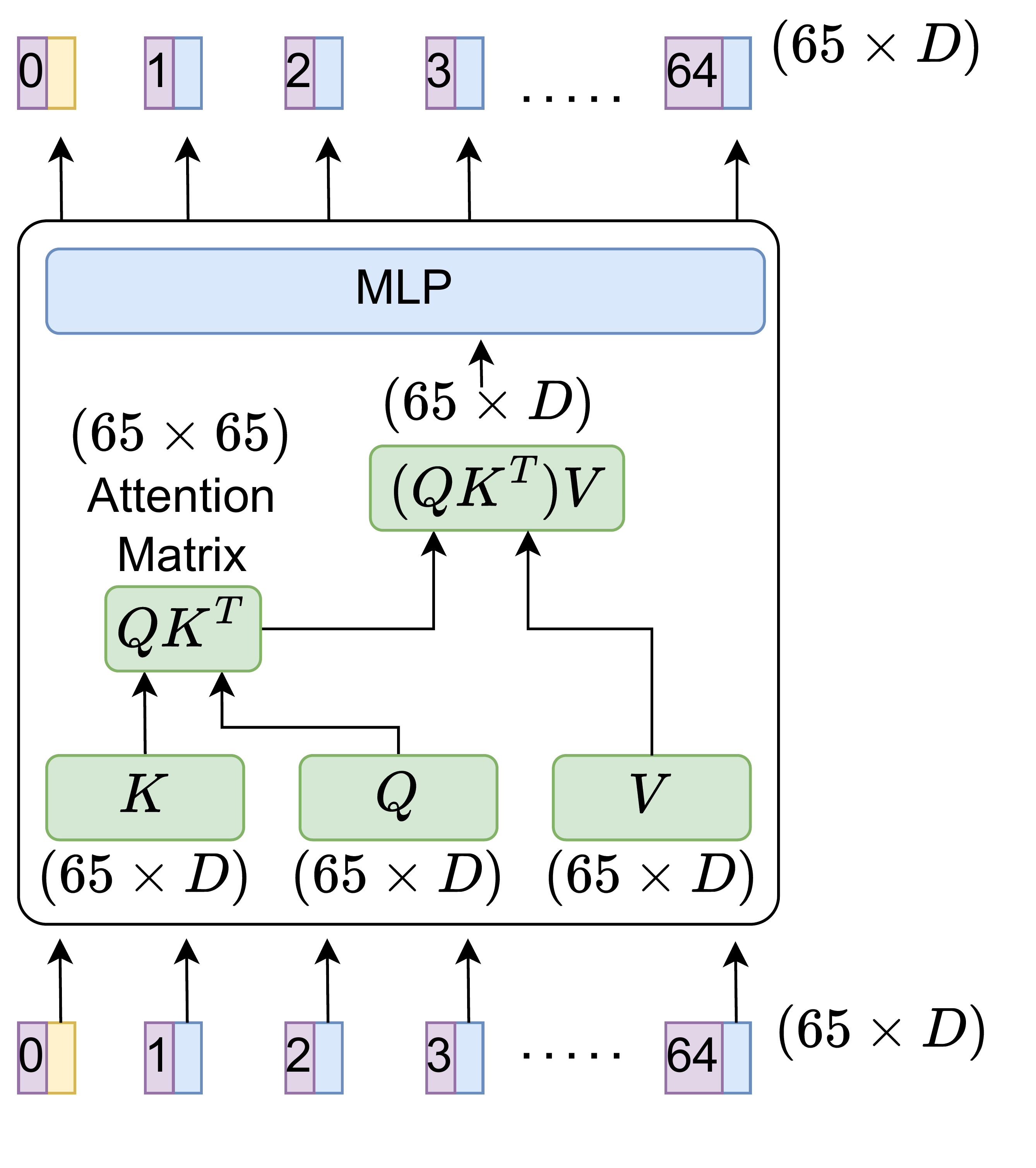}
    \caption{Transformer block architecture, which maps a set of input tokens $\mathbb{R}^{65\times D}$ into a similarly sized output. This block is used in the larger proposed architecture in \cref{fig:transformer_full}} 
    \label{fig:transformer_block}
\end{figure}

\begin{table}

    \begin{tabular}{|P{0.58in}|P{0.49in}|P{0.38in}|P{0.36in}|P{0.62in}|}\hline
        \rowcolor[HTML]{C0C0C0} 
        \textbf{Model} & \textbf{Hidden Dim ($D$)} & \textbf{Layers ($L$)} & \textbf{Heads ($h$)} & \textbf{Parameters (approx.)} \\
        \hline
        ViT-Tiny  & 192              & 12           & 3           & 5M                   \\\hline
        ViT-Small & 384              & 12           & 6           & 22M                  \\\hline
        ViT-Base  & 768              & 12           & 12          & 86M                  \\\hline
        ViT-Large & 1024             & 24           & 16          & 307M                 \\\hline
        ViT-Huge  & 1280             & 32           & 16          & 632M                
    \\ \hline\end{tabular}%

    \caption{Comparison of ViT sizes, including hidden dimension $D$, number of layers $L$, number of heads $h$, and total number of trainable parameters.}
    \label{table:ViTs}
\end{table}

\subsection{Vision Transformers}\label{sec:3a}
Vision Transformer (ViT) \cite{dosovitskiy2020image} is a deep learning architecture that leverages self-attention based transformer blocks for image processing, rather than natural language processing. ViTs rely on the patchify and position embedding process to map images into token sequences for transformer processing.

ViTs come in different configurations that vary primarily in three hyperparameters: the hidden dimension size $D$, the number of encoder layers $L$, and the number of attention heads $h$. These parameters scale the model’s representational capacity and computational cost. Smaller variants ("Tiny" or "Small") use lower hidden dimensions, fewer heads while keeping moderate depth. The larger ViTs increase the scale of both the depth and hidden dimension to capture more complex relationships at the expense of memory and runtime.  A summary comparison of these model families is provided in \cref{table:ViTs} \cite{dosovitskiy2020image, zhai2022scalingvisiontransformers}.

Larger architectures rely on longer training and more extensive datasets to achieve higher performance. For small datasets, the representative capacity of the model becomes less important, and the number of parameters can introduce too much complexity. Smaller models might show better performance on small datasets, as fewer samples are available for the architecture to extract relationships from. Large datasets, however, have more capacity to train larger models as there are enough samples for complex patterns to be recognized. 

Beyond architectural scaling, an important training strategy for improving representation learning in vision transformers involves token masking. In this approach, during training, a random subset $\mathcal{M} \subset \{1,\dots,N\}$ of patch tokens from an image sequence is selected for masking. Tokens corresponding to indices in $\mathcal{M}$ are replaced by a special learned embedding $\mathbf{z}_{\text{mask}}$, while unmasked tokens remain unchanged

\begin{equation}
\tilde{\mathbf{z}}_i = 
\begin{cases} 
    \mathbf{z}_i & i \notin \mathcal{M} \\ 
    \mathbf{z}_{\text{mask}} & i \in \mathcal{M} 
\end{cases}
\end{equation}

The network’s objective is then to reconstruct the original embeddings or pixel values corresponding to these masked patches using contextual information from unmasked tokens. This forces the model to learn meaningful global representations rather than relying solely on local cues. The reconstruction loss is typically formulated as mean squared error (MSE) over only the masked positions

\begin{equation}
\mathcal{L}_{\text{mask}} = \frac{1}{|\mathcal{M}|} \sum_{i \in \mathcal{M}} \|\hat{\mathbf{x}}_i - \mathbf{x}_i\|^2
\end{equation}
where $\hat{\mathbf{x}}_i$ denotes the reconstructed patch embedding and $\mathbf{x}_i$ is the ground truth embedding for patch $i$. This masking-based pretraining strategy has been shown to significantly improve downstream performance by encouraging robust feature extraction from incomplete visual information \cite{dosovitskiy2020image}. This strategy also encourages a reliance on multiple tokens across the image, preventing the model from overfitting on small subsets of the image domain.

\section{Methodology}\label{sec:4}

\begin{figure*}
    \centering
    \includegraphics[trim={5mm 145mm 55mm 45mm}, clip, width=1\linewidth]{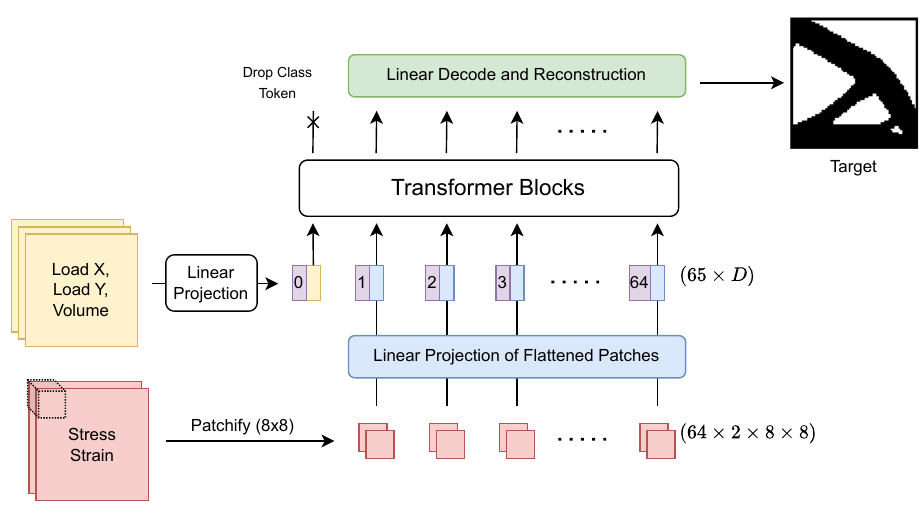}
    \caption{Full proposed transformer architecture, including a separate treatment of the loading conditions and volume fraction from the stress and strain fields.}
    \label{fig:transformer_full}
\end{figure*}

\subsection{Dataset}\label{sec:4a}
The dataset used in this study consists of structural topology representations paired with physical quantities that serve as surrogates for the sample's boundary conditions. Each sample topology is represented as a binary grayscale image of the design domain, where material and void regions are represented as 1 and 0, respectively. This representation also reflects the design densities after convergence from the topology optimization method, with a Heaviside filter to ensure binary output images. Associated with each topology are the boundary conditions and load, encoded as fields of the physical quantities strain energy density and von Mises stress. These fields are calculated on the initial design domain and not on the final topology. This ensures that the fields can be calculated prior to determining the optimal topology, so that they may be used as inputs for the model. These fields are normalized to a fixed intensity range from 0 to 1. In addition to these 2D representations, a volume fraction scalar is also associated with each sample.

Each sample is generated from a $64 \times 64$ domain of square elements. The load is randomly sampled from an element along the boundary of the domain and has a unit magnitude. The direction of the load is sampled from an angle of $0$ to $360^\circ$ from 6 evenly distributed angles. Each sample has a number of boundary conditions chosen between 1 and 4, from a set of 16 possible conditions. Nodes on the corners and face mid-points are possible boundary conditions. The sets of nodes between corners and midpoints are also boundary conditions, producing 8 total node sets and 8 node points. Nodes are fixed in both the x and y directions. The volume fraction is also a randomly sampled quantity, ranging from 30\% to 50\%.

The von Mises stress and strain energy density are evaluated at the element centers of the initial unoptimized domain during the pre-processing of the dataset. The finite element analysis assumes a unit linear elastic modulus and a Poisson's ratio of 0.3. These values only determine the scaling of the fields, rather than their relative distributions. In total, 60,000 samples were generated, with further dataset diversity being achieved from rotating and mirroring the samples.

\subsection{Architecture}\label{sec:4b}
The proposed model architecture is based on the ViT framework for vision transformer models. The input of the model is an image of the strain energy density and von Mises stress. This input is a two-channel input, as opposed to the three channels of image-based models. To divide the input domain into tokens, $8 \times 8$ patches are used to split the $64 \times 64$  domain into $64$ total segments. These segments are used as the input tokens to the transformer architecture for further processing. \cref{fig:patches} depicts this first step, demonstrating how the local patches of each image are kept together. The load direction, magnitude, and boundary conditions are fed into a NN model and output into the size of a single token in the transformer token domain. This allows the information contained in the sparse fields to be combined with the information in the dense strain energy and stress fields. To prevent the model from over-reliance on small regions of the input domain, we implement token masking on the stress and strain input fields.

These tokens are then used to compute the query $Q$, key $K$, and value $V$ matrices of the transformer's self-attention model. The result of this block is fed into a multi-layer perceptron. After multiple transformer blocks are used, the class token is dropped, and the information in the tokens is reconstructed into a single-channel (black and white) image. This image, with normalized values, can be interpreted as a density map across the design domain. A single forward pass of the transformer architecture will map the stress and strain field information into a density map, conditioned by the volume fraction, load conditions, and boundary conditions. The loss of the model can then be formulated as the pixel-to-pixel difference between the ground truth topology and the transformer model output.

\cref{fig:transformer_full} depicts the architecture for the transformer and the path that the data travels through various transformations. The input from the stress and strain fields is taken as an image input into the model, while the load and volume fraction are projected into an additional class token.

\subsection{Auxiliary Losses}\label{sec:4c}
To maximize the realism and manufacturability of the design, the model is also directly penalized on the performance of a few derivable evaluation metrics. These auxiliary losses adjust the trajectory of the gradient descent algorithm toward models that more reliably produce these desirable traits. The actual volume fraction of the model output can be directly computed and compared with the input target volume fraction, giving a loss function $VF(\tilde \rho)$. This loss term is a function of the transformer's predicted density field $\tilde \rho$. The load discrepancy $L(\tilde \rho)$ can be computed, which simply evaluates whether density is placed over the load condition. The other auxiliary loss used is the floating material loss $FM(\tilde\rho)$. These metrics ensure that the model is accurately reflecting the input to the model, rather than creating the most visually similar architecture that resembles the topologies common to the stress and strain field information. Without performing gradient descent over the compliance space, there is a risk that pixel space similarity may produce designs that have significantly different final compliance values. Performing a penalization on topology outputs that contain floating material can help alleviate this issue by encouraging more realistic designs. Designs that have floating bodies or two or more bodies for a single load condition are guaranteed to be worse (and are often substantially worse) than the optimal design for the requested input conditions. Penalizing the model for creating unrealistic and non-manufacturable designs can improve the performance by providing simple, easy to learn relationships for the model to optimize. The equations for the auxiliary losses are as follows

\begin{figure}
        \centering
    \includegraphics[width=1\linewidth]{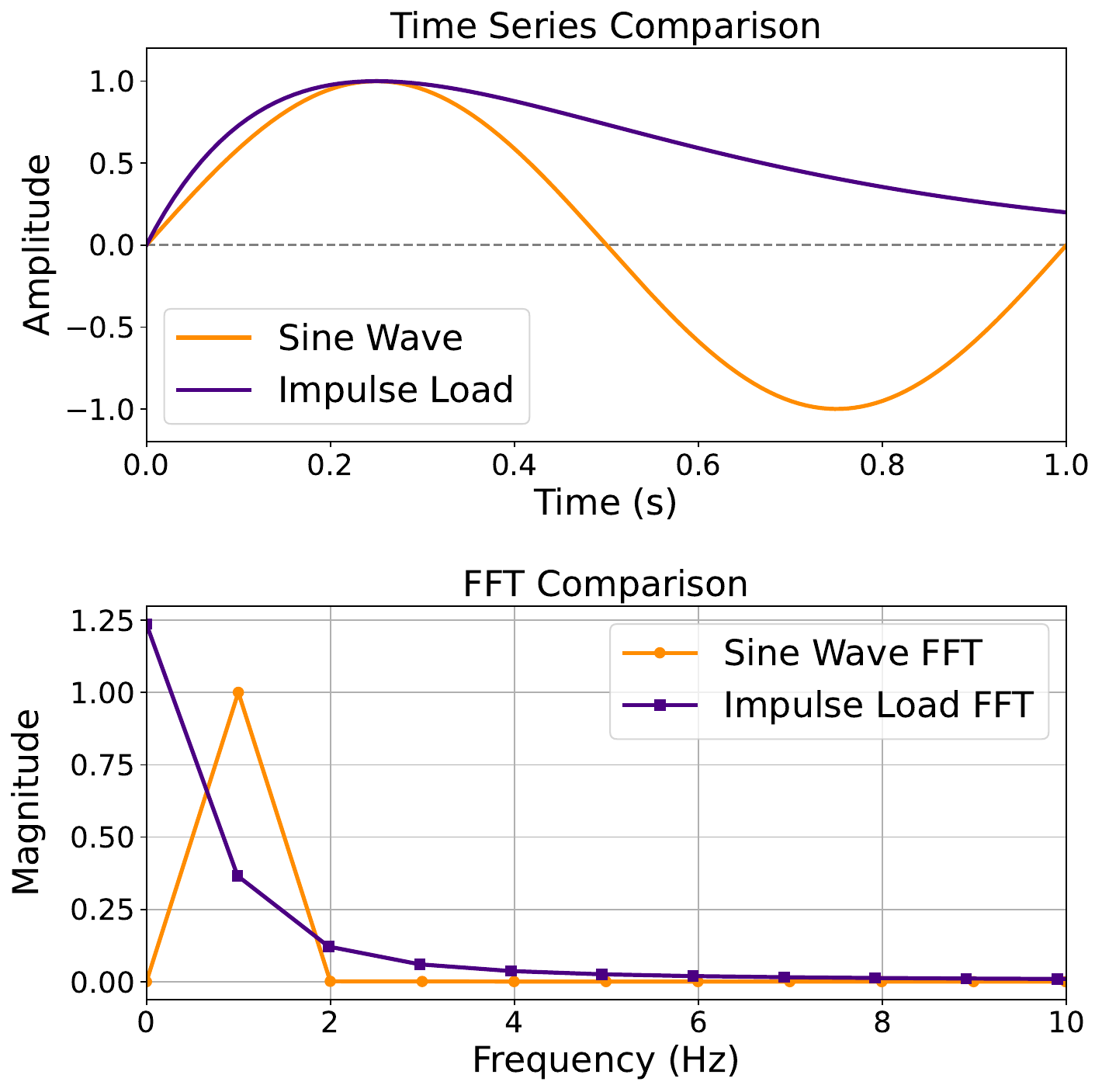}
    \caption{Plot of the time sequence and FFT amplitude transformation for the two loads used in the dynamic dataset, sine (orange) and impulse (indigo).} 
    \label{fig:loads}
\end{figure}

\begin{eqnarray}
VF(\tilde{\rho}) = |f - V(\tilde{\rho})/V_0| \\
L(\tilde{\rho}) = 1 - \sum_{e=1}^{N_e}\sqrt{(\tilde \rho_e F^x_e)^2 +  (\tilde \rho_eF^y_e)^2} \\
FM(\tilde{\rho}) = \begin{cases} 
    0 & k >1 \\ 
    1 & k = 1 
\end{cases}
\end{eqnarray}
where $V(\tilde \rho)$ is the volume of the design, $F_e$ is the load vector at at element $e$ with components in the $x$ and $y$ directions. $k$ is the number of components after implementing a connected components algorithm. We utilize the Python module Kornia \cite{eriba2019kornia} for determining connected components in a differentiable way. For models that consistently produce single structured designs that abide by the loading condition and volume fraction of the input, the auxiliary losses will quickly tend toward zero. These losses weigh more heavily in the early learning stages when the random parameters of the initialized model dominate the behavior.

\begin{table*}
\centering

\begin{tabular}{lllllll}
\rowcolor[HTML]{C0C0C0} 
\textbf{Metric}                    &  \textbf{TopoDiff}&\textbf{Tiny} & \textbf{Small} & \textbf{Base} & \textbf{Large}& \textbf{Huge} \\
Compliance Error (\%)              &                4.39&20.13& 1.88           & 9.31          & 10.13 & 5.07          \\
Compliance Error Above   30\% (\%) &                0.83&78.6& 5.2            & 21.4          & 17.8  & 5.0\\
Median Compliance Error   (\%)     &                2.56&5.26& 0.39           & 2.78          & 3.84  & 1.20\\
Volume Fraction Error (\%)         &                1.85&19.17& 1.01           & 4.75          & 2.19  & 1.39          \\
Load Discrepancy (\%)              &                0.0&60.4& 3.8            & 11.6          & 3.6   & 0.2\\
Floating Material (\%)             &                5.54&69.0& 12.2           & 84.8          & 85.6  & 48.6         
\end{tabular}%
\caption{Analysis metrics for the static topology optimization dataset with the ViT Tiny, Small, Base, Large, and Huge models.}
\label{table:static}
\end{table*}

\begin{figure}
        \centering
    \includegraphics[width=1\linewidth]{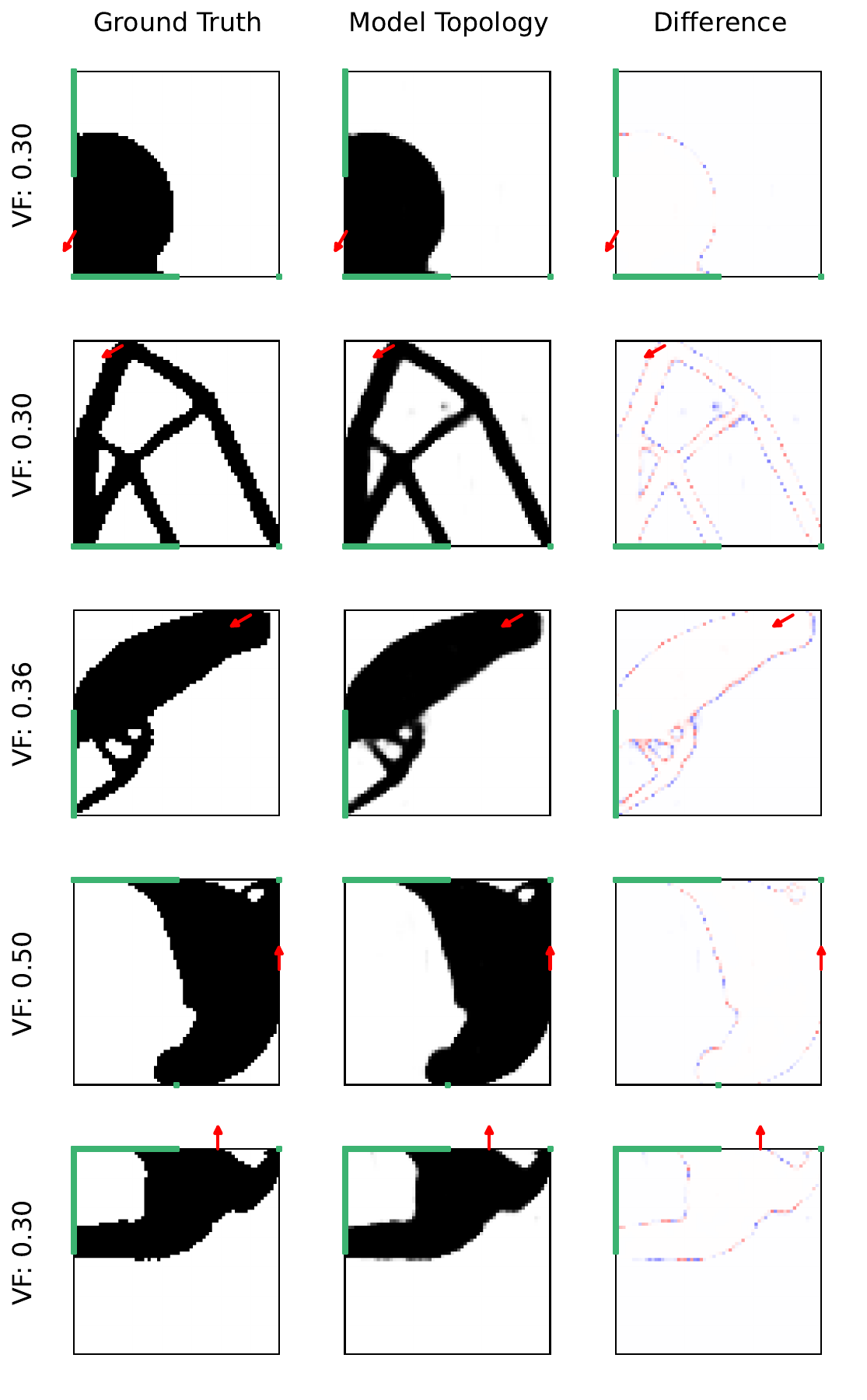}
    \caption{Validation samples from the ViT Small model. The left column shows the ground truth topology, the middle column shows the model's output, and the right column shows the difference between the two. Blue represents where the model applied material, where the ground truth didn't, while red represents the model missing a location to apply material.} 
    \label{fig:vitsmallsamples}
\end{figure}

\subsection{Transfer Learning on Dynamic Optimization}\label{sec:4d}
Dynamic topology optimization introduces additional complexity as the optimization process must account for the time-dependent response under various load and boundary conditions. As a result, running the dynamic optimization takes dramatically longer, making the data less available than the static optimization. The dataset available for the dynamic case has roughly 6,000 samples compared to the 60,000 in the static optimization dataset. Two time-dependent load types are utilized: sine and impulse. The magnitude functions of these loads and the FFT transformations for them are depicted in \cref{fig:loads}. The equation for the sine load $s(t)$ is \cref{eqn:sine} and the equation for the impulse load $i(t)$ is \cref{eqn:impulse}, both as functions of time $t$. For simplicity, we consider the range from $0$ to $1$ seconds. 

\begin{align}
    s(t) = \sin\!\left( 2 \pi t \right) \label{eqn:sine} \\
    i(t) = \frac{t}{0.25} e^{-\tfrac{t}{0.25} + 1} \label{eqn:impulse}
\end{align}

To accommodate this smaller dataset size, we implement a transfer learning framework. We expect many of the low and mid-level patterns to be similar between the static and dynamic cases. By using the transformer model pre-trained on the static dataset, we are able to leverage the larger dataset to tune most of our model parameters, leaving a fine-tuning step for the smaller dynamic dataset. This reduces the computational burden of training from scratch and reduces the risk of overfitting to the smaller dynamic dataset.

Compared to the static optimization, the dynamic case has more complex load information. Rather than a single magnitude, the input to the model architecture should incorporate the time-based response. To encode this information into the token, we reconfigure the class token projection to also take the first 10 frequencies from the FFT transform of the load magnitude functions. This allows the class projection layer to incorporate the new dynamic information into the token sequence.

\begin{table*}
\label{table:dynamic}
\begin{tabular}{lllll}
\rowcolor[HTML]{C0C0C0} 
\textbf{Metric} & \textbf{Untrained} & \textbf{Class Projection} & \textbf{Decoder Projection} & \textbf{Decoder Layers} \\
Compliance Error (\%)              & 15.14 & 12.72 & 12.93 & 4.81 \\
Compliance Error Above   30\% (\%) & 21.8  & 44.4  & 35.6  & 16.6 \\
Median Compliance Error   (\%)     & 4.18  & 4.00& 3.74  & 0.22 \\
Volume Fraction Error (\%)         & 1.62  & 6.39  & 7.49  & 4.38 \\
Load Discrepancy (\%)              & 5.00& 36.6  & 22.2  & 9.20\\
Floating Material (\%)             & 21.4  & 42.6  & 28.2  & 48.0\end{tabular}%
\caption{Analysis metrics for the dynamic topology optimization dataset on the transfer learning model, using ViT-Small as the pre-trained model. The untrained model is compared against the class projection tuning, the decoder projection tuning, and the decoder layers tuning models.}
\end{table*}

\section{Results and Analysis}\label{sec:5}
Each model is trained for 800000 iterations, with a batch size of 256. To assess the utility of the model, six metrics are used for analysis. Compliance error is the average relative difference in the compliance of the original optimized topology. Compliance error above 30\% is the proportion of samples that exhibit a compliance error above 30\%. These samples are usually easy to spot visually due to large discrepancies between the model topology and the ground truth. Median compliance error is also calculated, although this quantity is measured after removing the 'failed' samples with more than 30\% error.  The three auxiliary loss metrics, volume fraction error, load discrepancy, and floating material, are also used for assessing the models. Each model is sampled on a validation set of 500 samples to ensure that overfitting does not influence the results of the analysis. The same samples are used for all models to prevent random sampling skewing the results.

\begin{figure}
        \centering
    \includegraphics[width=1\linewidth]{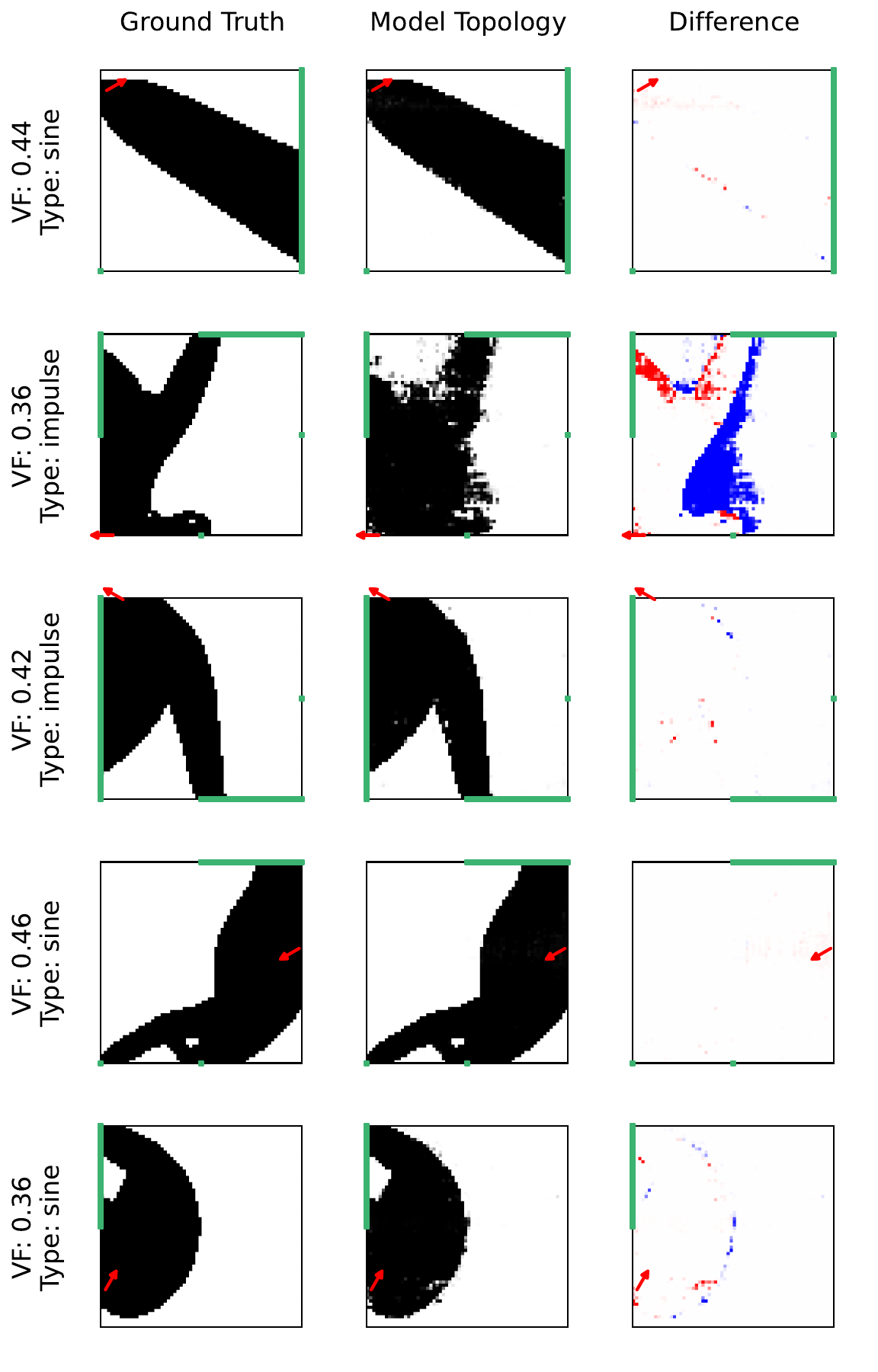}
    \caption{Validation samples from the ViT Small model finetuned on the decoder layers for dynamic data. The left column shows the ground truth topology, the middle column shows the model's output, and the right column shows the difference between the two.} 
    \label{fig:dyn_samples}
\end{figure}

For the static optimization, the five ViT model sizes (Tiny, Small, Base, Large, and Huge) are used for determining the best architecture. The results of the training and analysis for each of these models are shown in \cref{table:static} and compared against the current state-of-the-art model TopoDiff. The small model led to the lowest compliance error and lowest median compliance error, at 1.88\% and 0.39\% respectively. The larger models performed worse, likely due to the large number of parameters and relatively small dataset size compared to the datasets the ViT was originally designed for. This is also reflected in metrics such as the error rate above 30\%. Models larger than the Small architecture seem to produce disconnected structures, which leads to higher-than-expected overall compliance errors. The smallest ViT model, ViT-Tiny, also struggles with capturing relationships in the dataset, getting poor scores on even the easier metrics, like volume fraction error and load discrepancy. Samples from ViT-Small are shown in \cref{fig:vitsmallsamples}, depicting the ground truth topology, model topology, and difference.

For the dynamic optimization task, the best ViT architecture, ViT-Small, is used as the baseline model for transfer learning. First, we fine-tuned the ViT-Small with the altered class projection layer to include the dynamic FFT results. Next, we also trained the final projection layer in the decoder, where the tokens are projected into patches before the final reconstruction step. For the final architecture, we fine-tuned the two previous components as well as the final three layers of the transformer model. This provides a range of models in which different numbers of parameters are available for tuning. We refer to these three transfer learning models as "Class Projection," "Decoder Projection," and "Decoder Layers" respectively.

The results from the dynamic transfer learning case are presented in \cref{table:dynamic} alongside the untrained ViT-Small model, tested on the dynamic dataset. The models all struggled with the floating material category, indicating a lower performance than on the static optimization dataset. The highest performing model for our testing was the decoder layers model, which also had the highest number of trainable parameters. We achieved a 4.81\% compliance error on the dynamic dataset, although with a very high 48\% floating material error. This indicates a low confidence in the model, even though the topology performs relatively well after thresholding. A sample of the best model, the decoder layers model, is shown in \cref{fig:dyn_samples}. This figure also reflects the model's lack of confidence, as seen by blurry or noisy boundary edges.

Additionally, the runtime of the original topology optimization process was collected to determine what speedup would be expected from the transformer model approach. For static optimization, the topology optimization process finished in 66 seconds on average, while the transformer model was sampled in roughly 2 seconds. This represents a 33 times speedup for an average compliance error increase of 1.88\%. For the dynamic optimization, the optimization process took significantly longer, taking 8900 seconds on average. The sampling for the dynamic model took the same amount of time as the static optimization, 2 seconds, because the model architecture remains mostly the same between static and dynamic optimization. This results in a speedup of 4,450 times, significantly higher than the static optimization, at the cost of roughly 4.81\% increased compliance error.

\section{Conclusion}\label{sec:6}
In this work, we introduced a transformer model architecture with auxiliary losses as a novel approach to achieving speedups for structural topology optimization, extending the application to both static and dynamic formulations. Our architecture captures complex spatial relationships across the design domain by utilizing ViT models on an image input of stress and strain distributions calculated on the input conditions. Our experiments demonstrated that model size trades off accuracy and efficiency, while transfer learning between static and dynamic tasks improved convergence and predictive accuracy. Compared with classical iterative solvers, the trained models offered substantial inference-time efficiency gains, highlighting their potential for accelerating large-scale design problems, especially in the field of dynamic optimization.

\subsection*{Declarations}
\noindent \textbf{Conflict of interest} On behalf of all authors, the corresponding author states that there is no conflict of interest.

\noindent \textbf{Funding} This work has been financially supported by the Institute of Digital Engineering - USA.

\noindent \textbf{Author contributions} Conceptualization: Srijan Das, Alireza Tabarraei; Methodology: Aaron Lutheran; Formal analysis and investigation: Aaron Lutheran; Writing - original draft preparation: Aaron Lutheran; Writing - review and editing: Srijan Das, Alireza Tabarraei; Supervision, Srijan Das, Alireza Tabarraei.

\noindent \textbf{Ethics approval and Consent to participate} Not applicable for this work.

\noindent \textbf{Data Availability} Dataset will be made available on request to the corresponding author.

\noindent \textbf{Replication of results} Replication material, including model parameters, and code, are available on request to the corresponding author.

\nocite{*}
\bibliographystyle{elsarticle-num}
\bibliography{Transformer-Model.bib}

\end{document}